# Compensation of polarization dependent loss using noiseless amplification and attenuation

R.A. Brewster[1], B.T. Kirby[2], J.D. Franson[1], and M. Brodsky[2]
[1]*University of Maryland Baltimore County, Baltimore, Maryland 21250, USA*
[2]*United States Army Research Laboratory, Adelphi, Maryland 20783, USA*

Polarization dependent loss (PDL) is a serious problem that hinders the transfer of polarization qubits through quantum networks. Recently it has been shown that the detrimental effects of PDL on qubit fidelity can be compensated for with the introduction of an additional passive PDL element that rebalances the polarization modes of the transmitted qubit. This procedure works extremely well when the output of the system is postselected on photon detection. However, in cases where the qubit might be needed for further analysis this procedure introduces unwanted vacuum terms into the state. Here we present procedures for the compensation of the effects of PDL using noiseless amplification and attenuation. Each of these techniques introduces a heralding signal into the correction procedure that significantly reduces the vacuum terms in the final state. When detector inefficiency and dark counts are included in the analysis noiseless amplification remains superior, in terms of the fidelity of the final state, to both noiseless attenuation and passive PDL compensation for detector efficiencies greater than 40%.

## I. INTRODUCTION

The two main decoherence mechanisms affecting polarization photonic qubits transmitted through fiber optic networks are polarization mode dispersion (PMD) and polarization dependent loss (PDL) [1-3]. PDL, which is the attenuation of light as a function of polarization, introduces unavoidable loss and therefore its impact on transmitted qubits cannot be entirely rectified. While virtually nonexistent in modern optical fibers, PDL is present in nearly all network elements such as isolators, circulators, and amplifiers. Significant effort has been directed at understanding the impact of PDL in classical communication systems [4-8]. Recently this analysis has been expanded to entangled quantum systems [9-11]. These studies have mainly been concerned with understanding how the entanglement of a state is reduced by the presence of PDL [9-10], and with developing strategies for mitigating this [11].

In general, PDL reduces the overall probability that a state is transmitted through a channel, due to attenuation, as well as alters the states that are transmitted, due to its polarization dependence. Intuitively, PDL can be converted into pure loss through the introduction of additional PDL that is tuned such that the concatenation of the system PDL and the inserted PDL becomes polarization independent pure loss. While this strategy will recover the fidelity of the initial state upon postselection on photon detection, since there is no longer any polarization dependence in the system, it also introduces additional vacuum terms into the state.

In this paper we propose compensating for the effects of PDL on polarization encoded qubits by using noiseless amplification and attenuation and compare these with a previously proposed technique based on passive optical elements [10-11]. All three methods effectively convert the PDL of the system into a polarization independent net loss. The advantage of noiseless amplification and attenuation over additional passive PDL is that they, at least partially, herald that the correction has been successful. This heralding allows for a correction of the polarization modes of a transmitted qubit with fewer additional vacuum terms than in the passive case.

Throughout this paper we will model the effects of PDL as outlined in the first box on the left of fig. 1. Since loss can be modeled as a beam splitter coupled to the environment we treat PDL as two different beam splitters acting separately on the horizontal and vertical polarization modes of the photon. These beam splitters would have transmission factors $t_h$ and $t_v$ for the horizontal and vertical modes respectively. For simplicity, we let $t_v = 1$, thus we only consider loss in the horizontal mode.

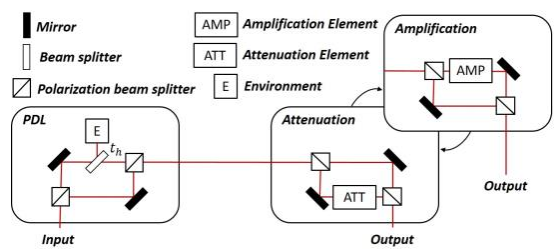

FIG. 1. Our input polarization qubit experiences unavoidable PDL modeled as shown in the first box labeled PDL. For simplicity, we consider the case where only the horizontal polarization mode was initially attenuated by a factor $t_h$. Following this PDL, we correct for it using either one of two possible attenuation schemes or an amplification scheme. These schemes are further detailed in Fig. 2. In the case of a correction using attenuation shown in the upper box, we attenuate the vertical mode to balance it with the horizontal mode. Similarly, in the case of amplification shown in the lower box, we amplify the horizontal mode to balance it with the vertical mode.

The consequences of PDL on polarization qubits is most directly seen by example. Consider the balanced input qubit

$$|\psi_0\rangle = \frac{1}{\sqrt{2}}(|H\rangle + |V\rangle), \qquad (1)$$

where $|H\rangle$ and $|V\rangle$ represent the horizontal and vertical modes of the photon respectively. After PDL, the output state becomes

$$|\psi\rangle = \frac{1}{\sqrt{2}}(\sqrt{t_h}|H\rangle + |V\rangle + \sqrt{1-t_h}|0\rangle), \qquad (2)$$

where $|0\rangle$ is the vacuum state corresponding to a photon being lost to the environment and we have neglected the ancillary beam splitter output mode. We see from eq. (2) that PDL has two corrupting effects on transmitted polarization qubits, the first is that the ratio of the polarization modes has changed, and the second is the introduction of vacuum terms. Therefore, we need a way to correct for PDL which can mitigate each of these sources of error.

This paper is structured as follows. In sec. II we discuss several different correction schemes and present the output state fidelity as compared to the input state and probability of success for each method. This will be done in the idealized case of perfect detectors. In sec. III we present a model for an imperfect detector and compute the fidelity of the output state as compared to the input state as a function of detector efficiency for each correction scheme. A summary and conclusion are presented in sec. IV.

## II. CORRECTING FOR PDL

Broadly speaking we will consider two different categories of methods for converting PDL into a polarization independent loss. The first is to insert additional attenuation into the system which is oriented orthogonal to the original such that the polarization dependence cancels. In our scenario pictured in fig. 1, where the horizontal mode is initially attenuated, this means further attenuating the vertical mode by an equivalent amount. We will consider both passive and noiseless attenuation. The second method of correction we will consider is to amplify the horizontal mode back to the point that the polarization dependence once again disappears. This is pictured as the scenario in fig. 1 labeled as 'Amplification.'

The use of passive attenuation as a method for correcting for the detrimental effects of PDL was recently explored, and even experimentally demonstrated, in several papers [10-12]. An example of a passive corrective element is pictured in the upper left box in fig. 2. To see how this works in our scenario consider the state in eq. (2) which has already been transmitted through a PDL element. By adding a passive attenuator that only acts on the vertical polarization mode with transmission $T$, the polarization modes of the state become

$$|\psi'\rangle = \frac{1}{\sqrt{2}}(\sqrt{t_h}|H\rangle + \sqrt{T}|V\rangle), \qquad (3)$$

where we have neglected normalization. By tuning $T$, such that $T = t_h$ we have

$$|\psi'\rangle = \sqrt{\frac{t_h}{2}}(|H\rangle + |V\rangle), \qquad (4)$$

which is the desired polarization qubit of eq. (1).

While passive attenuation successfully recovers the polarization qubit it requires either postselection on detection or involves the addition of vacuum terms into the state, which we have neglected in eqs. (3) and (4). This is problematic when subsequent quantum operations are technologically expensive and it is essential to maximize the probability of success of each gate. In fact, as we will see this correction scheme will actually lower the fidelity of the output state, when vacuum states are considered, more so than if we had not corrected at all.

The unwanted vacuum terms in the passive attenuation case are our motivation for considering both noiseless attenuation and amplification. Though these two techniques require additional elements, such as beamsplitters, detectors, or ancilla sources, they also allow for some amount of heralding on success and hence are able to reduce the vacuum term in the final state without postselection on a final detection.

In the case of noiseless attenuation, we again pass the vertical mode through a beam splitter with transmission $T$, however we now postselect on having no photons in the ancillary output mode. This form of attenuation was first introduced in ref. [13] and, as we will see, will give a better fidelity than that of the passive attenuation. The process of noiseless attenuation is shown in the lower-left box of fig. 2.

Finally, the process of noiseless amplification is outlined in the right box of fig. 2. This device is a component piece of the larger noiseless amplifier first introduced by T. C. Ralph and A. P. Lund [14]. When included in the polarization interferometer pictured in the 'Amplification' box of fig. 1 it is analogous to the polarization-qubit amplifier of ref. [15] with the exception that it only amplifies the horizontal mode and not the vertical mode. Noiseless amplification works using unbalanced teleportation and successfully amplifies the mode labeled 'in' in the 'Noiseless amplification' box of fig. 2 with the amplified state exiting the mode labeled 'out'. The entangled state of the teleportation process is created by passing a single ancilla photon labeled as $|1\rangle$ into the lower beam splitter in the 'Noiseless amplification' box of fig. 1 which has transmission $T$. One of the outputs of this beam splitter is then combined with the input state to be amplified at the 50-50 beam splitter pictured at the top of the 'Noiseless amplification' box in fig. 2. Finally, the process is successfully heralded when the states $\langle 0|$ and $\langle 1|$ are detected in the output modes of the 50-50 beam splitter.

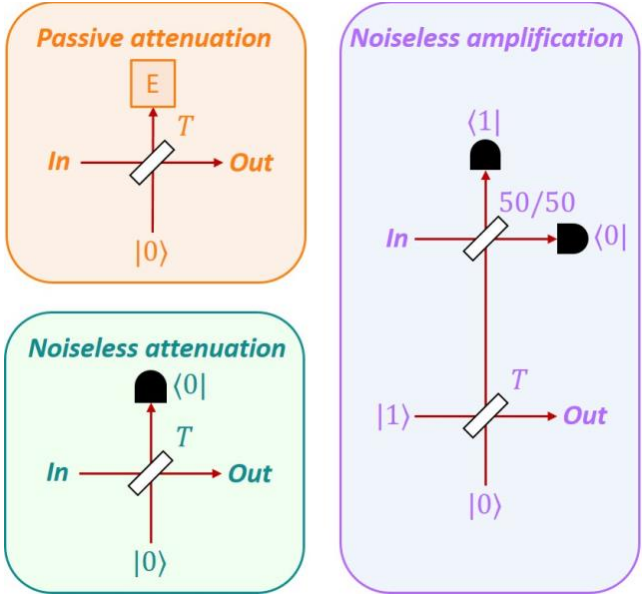

FIG. 2. A more detailed description of the three possible correction schemes. In passive attenuation the input state experiences loss where the ancillary mode is lost to the environment. In noiseless attenuation [Micuda] the same situation occurs except we instead postselect on vacuum in the ancillary mode. In noiseless amplification [Ralph AIP] an input state undergoes unbalanced teleportation by postselecting on the states specified at the detectors. As we will see, noiseless amplification provides for the best fidelity at the expense of transmission rate.

We will now calculate the output state of each of these three scenarios analytically. We do this for the general input qubit $|\psi_0\rangle$ given as

$$|\psi_0\rangle = c_1|H\rangle + c_2|V\rangle, \qquad (5)$$

where $c_1$ and $c_2$ are in general complex numbers satisfying

$$|c_1|^2 + |c_2|^2 = 1. \qquad (6)$$

This state corresponds to the input density operator $\rho_0$ as

$$\rho_0 = |\psi\rangle\langle\psi| = |c_1|^2|H\rangle\langle H| + c_2^*c_1|H\rangle\langle V| \\ + c_1^*c_2|V\rangle\langle H| + |c_2|^2|V\rangle\langle V|. \qquad (7)$$

We treat the evolution in figs. 1 and 2 using the unitary evolution operator for the beam splitter given as

$$U = \exp\left(i\arccos\left(\sqrt{T}\right)(a^\dagger b + ab^\dagger)\right), \qquad (8)$$

where $T$ is the transmission of the beam splitter and $a$ is the annihilation operator for one mode of the beam splitter, while $b$ is the annihilation operator for the other [16]. When postselecting, we apply an appropriate projection operator and any mode which is sent into the environment is traced out.

To begin, we calculate the state $\rho$ after undergoing an initial PDL, pictured in the left most box of fig. (1), which only attenuates the horizontal mode

$$\rho = |c_1|^2(1-t_h)|0\rangle\langle 0| + |c_1|^2 t_h |H\rangle\langle H| \\ + c_2^*c_1\sqrt{t_h}|H\rangle\langle V| + c_1^*c_2\sqrt{t_h}|V\rangle\langle H| \qquad (9) \\ + |c_2|^2|V\rangle\langle V|.$$

Inspection of eq. (9) reveals that there is now a vacuum term and even if we postselect on having a photon we still do not have the desired polarization state of equation (7).

We now consider the output state after correcting with additional PDL which has a transmission factor of $T$. The full density operator after tracing over the ancillary mode of the initial PDL and the unbalanced passive attenuation is $\rho_1$ given as

$$\rho_1 = \left[(1-t_h)|0\rangle\langle 0| + t_h(|c_1|^2|H\rangle\langle H| \\ + c_2^*c_1|H\rangle\langle V| + c_1^*c_2|V\rangle\langle H| \qquad (10) \\ + |c_2|^2|V\rangle\langle V|)\right],$$

where we have again let $T = t_h$ as was done in eq. (4). Note that if we postselect this state on detection [10-11] this would reduce to eq. (7) as expected.

The full output state for noiseless attenuation $\rho_2$, after postselecting on no photons at the detector shown in fig. 2, is given as

$$\rho_2 = \left[|c_1|^2(1-t_h)|0\rangle\langle 0| \\ + (|c_1|^2 t_h |H\rangle\langle H| + c_2^*c_1\sqrt{t_h T}|H\rangle\langle V| \qquad (11) \\ + c_1^*c_2\sqrt{t_h T}|V\rangle\langle H| + |c_2|^2 T|V\rangle\langle V|)\right].$$

If we let $T = t_h$ as we did before the state of eq. (11) becomes

$$\rho_2 = \Big[|c_1|^2(1-t_h)|0\rangle\langle 0|$$
$$+t_h(|c_1|^2|H\rangle\langle H|+c_2^*c_1|H\rangle\langle V| \qquad (12)$$
$$+c_1^*c_2|V\rangle\langle H|+|c_2|^2|V\rangle\langle V|)\Big].$$

Eq. (12) is not normalized because noiseless attenuation is a heralded process. The trace of eq. (12) gives the probability of success $P_2$ of the noiseless attenuation as

$$P_2 = t_h + |c_1|^2(1-t_h). \qquad (13)$$

Normalizing this probability of success away gives the output state conditioned on a success event as

$$\rho_2 = \frac{1}{t_h+|c_1|^2(1-t_h)}\Big[|c_1|^2(1-t_h)|0\rangle\langle 0|$$
$$+t_h(|c_1|^2|H\rangle\langle H|+c_2^*c_1|H\rangle\langle V| \qquad (14)$$
$$+c_1^*c_2|V\rangle\langle H|+|c_2|^2|V\rangle\langle V|)\Big],$$

which we see has a smaller vacuum state term than eq. (10).

Finally, for the noiseless amplifier the output state $\rho_3$ becomes

$$\rho_3 = \frac{1}{2}\Big[|c_1|^2(1-t_h)(1-T)|0\rangle\langle 0|$$
$$+(|c_1|^2 T t_h|H\rangle\langle H|+c_2^*c_1\sqrt{t_h T(1-T)}|H\rangle\langle V| \qquad (15)$$
$$+c_1^*c_2\sqrt{t_h T(1-T)}|V\rangle\langle H|$$
$$+|c_2|^2(1-T)|V\rangle\langle V|)\Big].$$

Since we are free to choose the parameter $T$, if we let

$$T = \frac{1}{1+t_h}, \qquad (16)$$

then equation (15) becomes

$$\rho_3 = \frac{t_h}{2(1+t_h)}\Big[|c_1|^2(1-t_h)|0\rangle\langle 0|$$
$$+(|c_1|^2|H\rangle\langle H|+c_2^*c_1|H\rangle\langle V| \qquad (17)$$
$$+c_1^*c_2|V\rangle\langle H|+|c_2|^2|V\rangle\langle V|)\Big].$$

From eq. (17), we see this choice of $T$ balances the qubit. Again, since noiseless amplification is a heralded process with probability of success $P_3$ given as

$$P_3 = \frac{t_h}{2(1+t_h)}\Big[1+|c_1|^2(1-t_h)\Big], \qquad (18)$$

the output state given a success event would be

$$\rho_3 = \frac{1}{1+|c_1|^2(1-t_h)}\Big[|c_1|^2(1-t_h)|0\rangle\langle 0|$$
$$+|c_1|^2|H\rangle\langle H|+c_2^*c_1|H\rangle\langle V| \qquad (19)$$
$$+c_1^*c_2|V\rangle\langle H|+|c_2|^2|V\rangle\langle V|\Big].$$

We can define the acceptance rate of any of the three PDL correction methods as the fraction of input states which are considered to be successfully prepared. For noiseless attenuation and amplification the acceptance rate is equivalent to the heralding probability. For passive attenuation the acceptance rate is unity since no information is given by the correction process about whether or not it was successful and hence there is no way to discriminate the output states. In fig. 3 we plot the acceptance rate of each process as a function of the magnitude of the initial PDL for an initial qubit with $c_1 = c_2 = 1/\sqrt{2}$. The PDL is expressed in decibels as [1]

$$\text{PDL [ dB ]} \equiv 10\log_{10}\left(\frac{t_{\max}}{t_{\min}}\right), \qquad (20)$$

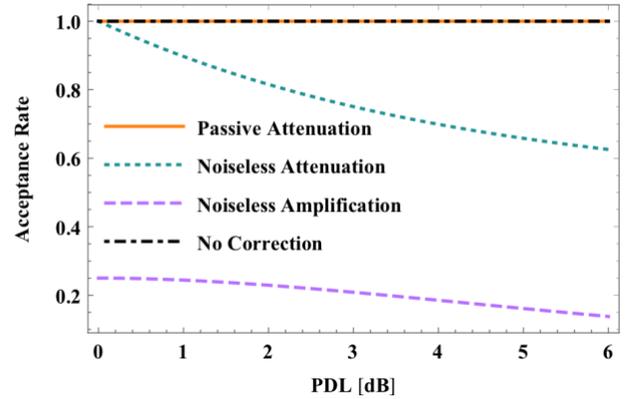

FIG. 3. Plot of the acceptance rate or probability of success of each of the possible correction schemes as a function of the PDL defined in eq. (20) in dB. Since passive attenuation is deterministic we will always accept the output state. From this figure we see that noiseless amplification will have a smaller acceptance rate than the other compensation schemes. In the noiseless amplification considered here we only accept one possible Bell state outcome. The acceptance rate for noiseless amplification might be able to be improved by accepting other Bell state outcomes and other techniques.

where, for the example considered here, $t_{\max}=1$ and $t_{\min}=t_h$. From fig. 3 we see noiseless amplification always has the lowest acceptance rate. As we will see this is the tradeoff needed to reduce the vacuum terms in the final state.

We now compare the acceptance rate to how close the output state is to the desired state. With the appropriate output states of eqs. (10), (14) and (19), we quantify the performance of each correction scheme by computing the fidelity given as [17]

$$F = \left[ \text{Tr}\left( \sqrt{\sqrt{\rho} \rho_0 \sqrt{\rho}} \right) \right]^2, \quad (21)$$

where $\rho$ is the relevant output density matrix and $\rho_0$ is the density matrix of the desire state given in eq. (7). The fidelity of the output states of each compensation scheme are plotted in fig. 4., for the case when $c_1 = c_2 = 1/\sqrt{2}$. Also in fig. 4 we plot the fidelity of the state after the initial PDL but before being corrected, given by eq. (9), for comparison.

Surprisingly, fig. 4 reveals that not correcting the state at all results in a better final state fidelity than either attenuation technique. This means that the vacuum terms added by the additional attenuation degrade the state fidelity more than it is improved by rebalancing the polarization modes. On the other hand, we see that noiseless amplification will always be the superior correction technique when output fidelity is the only concern. Comparing figs. 3 and 4 we see that there exists a tradeoff for noiseless amplification in that we can achieve a higher fidelity state after correction but at the expense of a low acceptance rate.

## III. CORRECTION WITH IMPERFECT DETECTORS

So far, we have only considered the case of ideal detectors. Since noiseless attenuation and noiseless amplification are heralded processes their performance may strongly depend on the efficiency of the detectors used. For this reason, we now examine how our calculations from section II change when detectors that are both inefficient and subject to dark counts are considered.

We model imperfect detectors as outlined in fig. 5 [18]. In this case the input state that we are attempting to detect is mixed with a thermal state $\rho_T$ at a beam splitter with transmission corresponding exactly to the detector efficiency $\eta$. We can define the thermal state $\rho_T$ as [19]

$$\rho_T = \frac{1}{1+\nu} \sum_{n=0}^{\infty} \left( \frac{\nu}{\nu+1} \right)^n |n\rangle\langle n|, \quad (22)$$

where $\nu$ is the average number of photons in the thermal state. Using eq. (22) and vacuum input, $\nu$ can be related to the probability of measuring a dark-count photon per time step $P_d$ as [20]

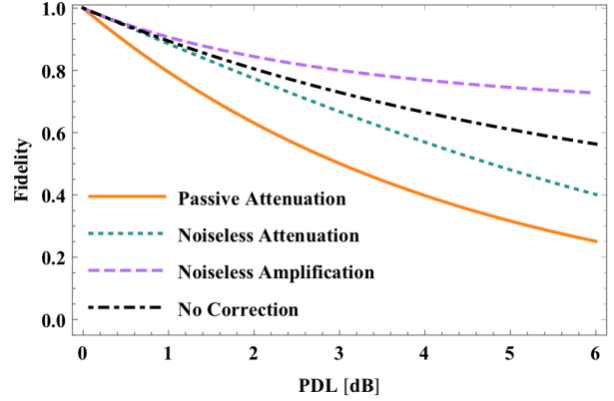

FIG. 4. Plot of the fidelity as a function of the PDL defined in eq. (21) in dB. As we can see, the fidelity will always be better in the case of noiseless amplification than for noiseless or passive attenuation. As mentioned in the text, this is due to the reduction in the probability amplitude of the vacuum term. This plot was generated for the ideal case of no detector noise.

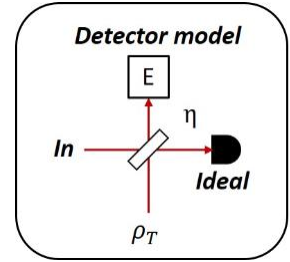

FIG. 5. Model used to simulate imperfect detectors with dark counts. An input state is mixed with a thermal state $\rho_T$ using an unbalanced beam splitter with transmission equal to the detector efficiency $\eta$. The temperature of the thermal state is chosen to be function of $\eta$ to guarantee a constant probability of finding a dark count photon per time step.

$$\nu = \frac{P_d}{(1-P_d)(1-\eta)}. \quad (23)$$

We again compute the new output states of all three corrective processes analytically. In the case of imperfect detectors, the output states are much more complicated and since inspection of the expressions themselves offers little physical insight we have moved them to the appendix. The results of a calculation of the fidelity are plotted in fig. 6 as a function of the detector efficiency $\eta$ for a constant PDL of 3dB and $P_d = 4 \times 10^{-5}$ photons per time step to coincide with the dark-count rate of the detectors of reference [21]. For reference, we have also included vertical lines in fig. 6 which indicate the efficiencies of real detectors. From fig. 6 we see that noiseless amplification remains superior for detector efficiencies greater than 40%.

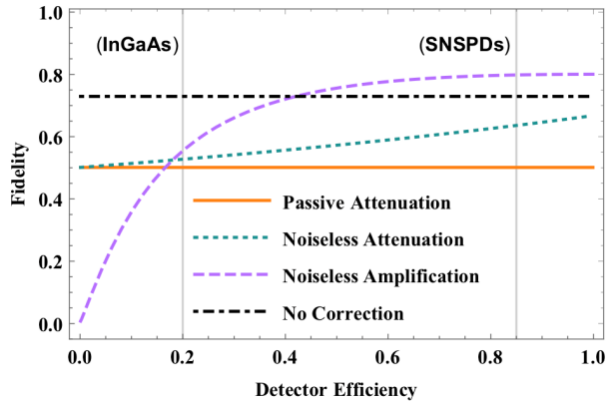

FIG. 6. Plot of the fidelity in the case of imperfect detectors as a function of detector efficiency $\eta$. We are modeling an imperfect detector according to the diagram in fig. 5. This plot was generated with an initial 3dB of PDL and a dark count rate of $4\times 10^{-5}$ photons per timestep. We see that in this case, noiseless amplification still does better for larger detector efficiencies. The vertical lines correspond to values for realistic detector efficiencies. The line at 20% represents Indium Gallium Arsenide (InGaAs) single photon detectors [21] and the line at 85% represents superconducting nanowire single-photon detectors (SNSPDs) [22].

In the limit as detector efficiency goes to unity the curves in fig. 6 approach the values reported in fig. 4 at 3dB. This makes sense as fig. 4 represents the case of ideal detectors. In the limit as detector efficiency goes to zero however we see that the fidelity due to the noiseless attenuation scheme approaches that of the passive attenuation scheme. This is because noiseless attenuation would be identical to passive attenuation in the absence of a detector. Finally, in the same limit the noiseless amplification scheme approaches a resulting fidelity of zero due to how heavily noiseless amplification is dependent on heralding.

## IV. SUMMARY AND CONCLUSIONS

We have considered the effects of polarization dependent loss under conditions where the output of a quantum communications system must always be accepted and used for further processing, as might be the case in a quantum repeater, for example. Polarization dependent loss and its passive compensation can reduce the overall fidelity under those conditions, since it introduces some probability amplitude for the vacuum state. Our results show that noiseless amplification gives a higher fidelity than either passive or noiseless attenuation under those conditions. Of course, there are situations in which the output is only accepted if the signal contains a photon, in which case passive attenuation [10-11] gives a higher acceptance rate than noiseless amplification.

We have seen in previous work that information lost to the environment can create a significant amount of decoherence in macroscopic quantum optical systems [23]. This is due to the introduction of which-path information. In the situation considered here, we see a similar phenomenon where losing information to the environment in the form of PDL can reduce the fidelity of single-photonic systems when the output must always be accepted. The fidelity can be improved using a noiseless amplifier, which does not leave any which-path information in the environment [24].

Although polarization dependent loss tends to be small in optical fibers, it can have a major effect in optical components such as isolators, circulators, and amplifiers. As a result, the techniques discussed here should be of practical importance in quantum communications systems.

## ACKNOWLEDGEMENTS

This work was supported in part by a Graduate Assistance in Areas of National Need (GAANN) Fellowship from the US Department of Education (Grant No. P200A150003) and by the NSF under Grant No. 1402708.

## APPENDIX

In this appendix we express the output density operators in the case of nonideal detectors. We do this for the noiseless attenuation and noiseless amplification schemes only as they are the only cases dependent on detector efficiency. For the case of noiseless attenuation, the unnormalized output density operator would be given as

$$\rho_2 = \frac{(2-T-t_h)(1+\nu-\eta\nu)-\eta(1-T)}{(1+\nu-\eta\nu)^2}|c_1|^2|0\rangle\langle 0| + \frac{t_h}{1+\nu-\eta\nu}|c_1|^2|H\rangle\langle H| \\ + \frac{T}{1+\nu-\eta\nu}|c_2|^2|V\rangle\langle V| + \frac{\sqrt{t_h T}}{1+\nu-\eta\nu}\left(c_2^*c_1|H\rangle\langle V| + c_1^*c_2|V\rangle\langle H|\right),$$

(A1)

where $\eta$ is the efficiency of the detector and $\nu$ is the average number of photons in the thermal state used to model dark counts given by eq. (22) of the main text. The parameter $\nu$ can be related to the probability of detecting a dark count photon using eq. (23). From eq. (A1) we see a choice of $T=t_h$ will rebalance the state. Note that this is the same choice for $T$ as the ideal case of noiseless attenuation. Using this choice of $T$ gives the output state as

$$\rho_2 = \frac{(1-t_h)(2+2\nu-2\nu\eta-\eta)}{(1+\nu-\eta\nu)^2}|c_1|^2|0\rangle\langle 0| + \frac{t_h}{1+\nu-\eta\nu}(|c_1|^2|H\rangle\langle H|+c_2^*c_1|H\rangle\langle V| \quad \text{(A2)}$$
$$+c_1^*c_2|V\rangle\langle H|+|c_2|^2|V\rangle\langle V|).$$

Normalizing equation (A2) and inserting it into eq. (21) of the main text gives the resulting state fidelity after the noiseless attenuation scheme with nonideal detectors. This fidelity is plotted as the green dotted curve in fig. 6 for the case of 3dB of initial PDL.

In the case of noiseless amplification, the output state using two identical detectors with efficiency $\eta$ and average number $\nu$ would be

$$\rho_3 = \frac{1-T}{4\eta(1+\nu-\eta\nu)^2}\left[2(1-t_h)\left(\eta^2 + \frac{\eta\nu(1-\sqrt{\eta})^2(1-\eta)(1+\nu)}{(1+\nu-\eta\nu)^3} + \frac{\sqrt{\eta}\nu(1-\eta)(1+\nu+\eta\nu)}{(1+\nu-\eta\nu)^2}\right)\right.$$
$$+t_h(1-\eta)(1+\nu)\left(\eta\frac{(1-\sqrt{\eta})^2+(1-\eta)}{1+\nu-\eta\nu}+2\frac{\eta\nu(1-\sqrt{\eta})^2(1-\eta)(1+\nu)}{(1+\nu-\eta\nu)^4}+\frac{\sqrt{\eta}\nu(1-\eta)(1+\nu+\eta\nu)}{(1+\nu-\eta\nu)^3}\right)\bigg]|c_1|^2|0\rangle\langle 0|$$
$$+\frac{T(\nu(2-t_h)(1-\eta)+t_h)}{2(1+\nu-\eta\nu)^3}|c_1|^2|H\rangle\langle H|+\frac{T\nu(1-\eta)^2(1+\nu)}{2(1+\nu-\eta\nu)^4}|c_1|^2|2\rangle\langle 2|$$
$$+\frac{1-T}{2\eta(1+\nu-\eta\nu)^2}\left(\eta^2+\frac{\eta\nu(1-\sqrt{\eta})^2(1-\eta)(1+\nu)}{(1+\nu-\eta\nu)^3}+\frac{\sqrt{\eta}\nu(1-\eta)(1+\nu+\eta\nu)}{(1+\nu-\eta\nu)^2}\right)|c_2|^2|V\rangle\langle V| \quad \text{(A3)}$$
$$+\sqrt{(1-T)Tt_h}\left[\frac{\left(\sqrt{\eta}+(1+(2+i)\sqrt{\eta}-2\eta-(2+i)\eta^{3/2}+\eta^2)\nu+\left(1+(1+i)\sqrt{\eta}\right)(1-\eta^2)\nu^2\right)}{2(1+\nu-\eta\nu)^4}c_2^*c_1|H\rangle\langle V|\right.$$
$$+\frac{\left(\sqrt{\eta}+(1+(2-i)\sqrt{\eta}-2\eta-(2-i)\eta^{3/2}+\eta^2)\nu+\left(1+(1-i)\sqrt{\eta}\right)(1-\eta^2)\nu^2\right)}{2(1+\nu-\eta\nu)^4}c_1^*c_2|V\rangle\langle H|\bigg],$$

where the state $|2\rangle$ corresponds to the amplifier having erroneously added an extra horizontally polarized photon giving a state with one horizontal and one vertical photon. While not as obvious as the case of noiseless attenuation, we can still set the $|H\rangle\langle V|$ and $|V\rangle\langle H|$ terms equal to one other to find the optimal value for $T$. Using this optimal value for $T$ in eq. (A3) and normalizing we could then insert this state into eq. (21) of the main text to get the state fidelity. This state fidelity is plotted as the blue dashed line in fig. 6.